\documentclass[12pt]{article}

\newcommand{\be}{\begin{equation}}
\newcommand{\ee}{\end{equation}}
\newcommand{\bi}[1]{\vspace{-3mm} \bibitem{#1}}
\usepackage{epsfig,amsmath,amssymb,graphics,graphicx}

\begin{document}

\begin{center}
{\it Annals of Physics 323 (2008) 2756-2778}
\vskip 5 mm

{\Large \bf Fractional Vector Calculus and

\vskip 3mm

Fractional Maxwell's Equations}
\vskip 7 mm

{\large \bf Vasily E. Tarasov} \\

\vskip 3mm

{\it Skobeltsyn Institute of Nuclear Physics, \\
Moscow State University, Moscow 119991, Russia } \\
{E-mail: tarasov@theory.sinp.msu.ru}
\end{center}

\vskip 3 mm

\begin{abstract}
The theory of derivatives and integrals of non-integer order goes back to 
Leibniz, Liouville, Grunwald, Letnikov and Riemann.
The history of fractional vector calculus (FVC) has only 10 years.
The main approaches to formulate a FVC, which are used in the physics during 
the past few years, will be briefly described in this paper.
We solve some problems of consistent formulations of FVC 
by using a fractional generalization of the Fundamental Theorem of Calculus.
We define the differential and integral vector operations.
The fractional Green's, Stokes' and Gauss's theorems are formulated. 
The proofs of these theorems are realized for simplest regions.
A fractional generalization of exterior differential calculus
of differential forms is discussed.
Fractional nonlocal Maxwell's equations and
the corresponding fractional wave equations are considered.
\end{abstract}

\vskip 3 mm

\noindent
PACS: 45.10.Hj; 03.50.De; 41.20.-q\\


\vskip 7 mm

\newpage
\section{Introduction}

The fractional calculus has a long history 
from 30 September 1695, when the derivative of 
order $\alpha=1/2$ has been described by Leibniz \cite{OS,SKM} (see also \cite{Ross}).
The theory of derivatives and integrals of non-integer order goes back to 
Leibniz, Liouville, Grunwald, Letnikov and Riemann.
There are many interesting books about fractional calculus and 
fractional differential equations
\cite{OS,SKM,MR,Podlubny,KST} (see also \cite{K,N}).
Derivatives and integrals of fractional order, and 
fractional integro-differential equations 
have found many applications in recent studies in physics
(for example, see books \cite{Zaslavsky1,GM,WBG,H}, 
and reviews \cite{Zaslavsky2,MS,MK}).

The history of fractional vector calculus (FVC) is not so long. 
It has only 10 years and can be reduced to the papers \cite{BA1}-\cite{T3}.
The main approaches to formulate a FVC, which are used in the physics during 
the past few years, will be briefly described in this paper.
There are some fundamental problems of consistent formulations of FVC 
that can be solved by using a 
fractional generalization of the Fundamental Theorem of Calculus.
Fractional vector calculus is very important to describe 
processes in fractal media (see for example \cite{GM}).
A consistent FVC can be used in fractional electrodynamics \cite{E1,E2,E3,T1} 
and fractional hydrodynamics \cite{MMW,Media}.

In Section 2, we describe different approaches to formulate FVC, 
which are used in the physics during the past 10 years.
The problems of consistent formulation of FVC are described in Section 3.
A fractional generalization of the Fundamental Theorem of Calculus
is considered in Section 4.
In Section 5, the differential and integral vector operations are defined.
In Sections 6-8, the fractional Green's, Stokes' and Gauss's theorems are formulated. 
The proofs of these theorems are realized for simplest regions.
In Section 9, a fractional generalization of exterior calculus
of differential forms is discussed.
In Section 10, fractional nonlocal Maxwell's equations and
the corresponding fractional wave equations are considered.

\section{Approaches to fractional vector calculus}

For Cartesian coordinates, fractional generalizations of 
the divergence or gradient operators can be defined by
\be
grad^{\alpha} \, f(x)= {\bf e}_s D^{\alpha}_s f(x) ,
\ee
\be
div^{\alpha} \, {\bf F}(x) = D^{\alpha}_s F_s(x) ,
\ee
where $D^{\alpha}_s$ are fractional (Liouville, Riemann-Liouville, Caputo, etc.) 
derivatives \cite{OS,SKM,MR,Podlubny,KST} of order $\alpha$ 
with respect to $x_s$, ($s=1,2,3$).
Here ${\bf e}_s$ $(s=1,2,3)$ are orthogonal unit vectors, and
$F_s(x)$ are components of the vector field
\be \label{bfF} {\bf F}(x)=F_s(x) {\bf e}_s=F_x {\bf e}_x+F_y {\bf e}_y+F_z {\bf e}_z .
\ee

The main problem of formulation of FVC appears, when we try to 
generalize the curl operator and the integral theorems.
In Cartesian coordinates, the usual (integer) curl operator 
for the vector field (\ref{bfF}) is defined by
\be \label{curl}
curl {\bf F}= {\bf e}_l \varepsilon_{lmn} D_m F_n ,
\ee
where $D_m=\partial / \partial x_m$, and 
$\varepsilon_{lmn}$ is Levi-Civita symbol, which 
is $1$ if $ (i, j, k)$ is an even permutation of $(1,2,3)$,  
$(-1)$ if it is an odd permutation, and $0$ if any index is repeated.
The Fourier transform of the curl operator is 
\be
{\cal F} \left( curl {\bf F}(x) \right)= 
 {\bf e}_l \varepsilon_{lmn} (i k_m) \tilde F_n(k) ,
\ee
where
\be \label{Fk}
\tilde F_n(k)= {\cal F} \left( F_n(x) \right)=  
\int^{+\infty}_{-\infty} d^3x \, e^{-ikx} F_n(x) .
\ee
To define a generalization of (\ref{curl}), we can use a fractional
integro-differentiation instead of the derivative $D_m$.

\subsection{Ben Adda's fractional vector calculus}

In the paper \cite{BA1} (see also \cite{BA2}), fractional generalizations 
of gradient, divergence and curl operator for analytic functions
have been suggested in the form
\be
grad^{\alpha} \, f(x)= \frac{1}{\Gamma(\alpha+1)}{\bf e}_s D^{\alpha}_s f(x) ,
\ee
\be
div^{\alpha} \, {\bf F}(x) = \frac{1}{\Gamma(\alpha+1)} D^{\alpha}_s F_s(x) ,
\ee
\be 
curl^{\alpha} \, {\bf F}= \frac{1}{\Gamma(\alpha+1)} 
{\bf e}_l \varepsilon_{lmn} D^{\alpha}_m F_n (x) ,
\ee
where $\Gamma(\alpha+1)$ is the Gamma function.
In these definitions, the Nishimoto fractional derivative \cite{N} 
(see also Section 22 of \cite{SKM}) is used. 
This derivative is a generalization of the Cauchy's differentiation formula.
 
Fractional generalizations of integral operations (flux and circulation), 
and generalizations of Gauss's, Stokes', Green's integral theorems are not considered.

\subsection{Engheta's fractional vector calculus}

In the paper \cite{E1} (see also \cite{E1,E2,E3}),  
a fractional generalization of curl operator
has been suggested in the form
\be \label{ENcurl}
curl^{\alpha} \, {\bf F}= {\bf e}_l \varepsilon_{lmn} D^{\alpha}_m F_n (x) ,
\ee
where $D^{\alpha}_m$ are fractional Liouville derivatives \cite{KST}
of order $\alpha$ with respect to $x_m$, ($m=1,2,3$), that are defined by
\be \label{LD}
D^{\alpha}_{m} f(x):=
\lim_{a\rightarrow -\infty} \,  _aD^{\alpha}_{x_m} f(x_m) .
\ee
Here $ _aD^{\alpha}_{x}$ is the Riemann-Liouville derivative
\be
_aD^{\alpha}_{x} f(x)=\frac{1}{\Gamma(n-\alpha)} \frac{\partial^n}{\partial x^n}
\int^{x}_{a} \frac{f(x')}{(x-x')^{\alpha-n+1}} dx' , \quad (n-1 <\alpha<n) .
\ee
The fractional Liouville derivative (\ref{LD}) 
can be defined through the Fourier transform by
\be
D^{\alpha}_m F_n (x)=
{\cal F}^{-1} \left( (i k_m)^{\alpha} \tilde F_n(k) \right)=
\frac{1}{(2\pi)^3} \int^{+\infty}_{-\infty} d^3k \, e^{ikx} 
(i k_m)^{\alpha} \, \tilde F_n(k) ,
\ee
where $\tilde F_n(k)$ is defined by (\ref{Fk}) and 
$i^{\alpha}=\exp \{i \alpha \pi \, sgn(k)/2\}$. 
For this fractional curl operator, 
the fractional integral Stokes' and Green's theorems are not suggested.
The problems of a generalization of these theorems 
will be considered in the next section.

In general, the fractional vector calculus must include generalizations of
the differential operations (gradient, divergence, curl), 
the integral operations (flux, circulation), and the theorems of Gauss, Stokes and Green.

\subsection{Meerschaert-Mortensen-Wheatcraft fractional vector calculus}

In the paper of Meerschaert, Mortensen and Wheatcraft \cite{MMW},
a fractional generalization of curl operator 
has been suggested as
\be \label{MMWcurl}
curl^{\alpha} {\bf F}= {\bf e}_l \varepsilon_{lmn} D_m I^{1-\alpha}_n F_n ,
\ee
where $I^{1-\alpha}_n$ are fractional integrals of order $(1-\alpha)$
with respect to $x_n$, ($n=1,2,3$).
Note that the integration $I^{1-\alpha}_n$ in (\ref{MMWcurl}) 
is considered with the index $n$ as the component $F_n$. 
The derivative $D_m=\partial/\partial x_m$ in Eq. (\ref{MMWcurl}) 
is considered with respect to $x_m$, where $m \ne n$.
Therefore expression (\ref{MMWcurl}) can be presented 
as the usual (integer) curl operator
\be \label{10}
curl^{\alpha} {\bf F}=curl {\bf F}(\alpha)
\ee
for the field 
\be \label{Fa}
{\bf F}(\alpha)={\bf e}_n I^{1-\alpha}_n F_n .
\ee
Equation (\ref{10}) allows us to use the usual (integer) 
integral Stokes' and Green's theorems.

In Eq. (\ref{MMWcurl}), the fractional integral $I^{\alpha}_n$ 
and the integer derivative $D_m$
have antisymmetric indices, and the components of (\ref{MMWcurl}) are
\be
\left( curl^{\alpha} {\bf F}\right)_{x}= 
D_y I^{1-\alpha}_z F_z - D_z I^{1-\alpha}_y F_y ,
\ee
\be
\left( curl^{\alpha} {\bf F}\right)_{y}= 
D_z I^{1-\alpha}_x F_x - D_x I^{1-\alpha}_z F_z ,
\ee
\be
\left( curl^{\alpha} {\bf F}\right)_{z}= 
D_x I^{1-\alpha}_y F_y - D_y I^{1-\alpha}_x F_x .
\ee
It is easy to see that operator (\ref{MMWcurl}) has no fractional derivatives
with respect to $x_m$, ($m=1,2,3$),
like as $D^{\alpha}_m = D_m I^{1-\alpha}_m$ or $^CD^{\alpha}_m= I^{1-\alpha}_m D_m$.

As a result, we have the usual (integer) vector calculus for 
new type of fields as in (\ref{Fa}).
The suggested approach cannot be considered as a fractional generalization
of vector calculus.
It is important to define a curl operator with fractional derivatives 
in such a form that fractional generalizations of the integral theorems exist.

\subsection{Other approaches to fractional vector calculus}

In the papers \cite{T1,Media}, fractional generalizations of 
integral operations and Gauss's, Stokes', Green's theorems
have been suggested. 
These generalizations are considered to describe fractional media 
by a continuous medium model. 
The differential operations are defined with respect to
fractional powers of coordinates. 
These operations are connected with 
fractional derivatives only by Fourier transforms (see \cite{TZ}).
As a result, an "ideal" fractional vector calculus is not suggested.

In the papers \cite{T2,T3}, fractional differential vector operations 
are considered by using fractional generalizations of differential forms
that are suggested in \cite{FDF} (see also \cite{YZH,Kaz}).  
A fractional gradient is defined by an exact fractional 1-form. 
A fractional curl operator is described by a fractional exterior derivative of
a fractional differential 1-form. 
The Riemann-Liouville  derivatives are used in \cite{T2}, and 
the fractional Caputo derivatives are used in \cite{T3}.
We have 
\be
grad^{\alpha} \, f(x)= {\bf e}_s \, _0^CD^{\alpha}_{x_s} f(x) ,
\ee
\be 
curl^{\alpha} {\bf F}= {\bf e}_l \varepsilon_{lmn}\, _0^C D^{\alpha}_{x_m} F_n ,
\ee
where $ _0^CD^{\alpha}_{x_m}$ is a fractional Caputo derivative with respect to $x_m$:
\be
_a^CD^{\alpha}_{x} f(x)=\frac{1}{\Gamma(n-\alpha)} 
\int^{x}_{a} \frac{1}{(x-x')^{\alpha-n+1}} dx' 
\frac{\partial^n f(x')}{\partial (x')^n}
, \quad (n-1 <\alpha<n) .
\ee
The fractional generalizations of integral theorems 
(Gauss's, Stokes', Green's theorems) are not considered
and the fractional integrals for differential forms are not defined.

\section{Problems of fractional generalization of Green's formula}

Let us describe a main problem that appears when the curl operator 
and integral formulas are generalized on a fractional case.
For simplification, we consider a rectangular domain on $\mathbb{R}^2$
and integral formulas in Cartesian coordinates.

The Green's formula in Cartesian coordinates is
\be \label{Green}
\int_{\partial W} \left( F_x dx +F_y dy \right)=
\int\int_{W} \, dx dy \left[ D_y F_x-D_x F_y \right] ,
\ee
where $F_x=F_x(x,y)$ and $F_y=F_y(x,y)$ are functions defined 
for all $(x,y)$ in the region $W$.

Let $W$ be the rectangular domain
\[ W:=\{(x,y) : \quad a\le x\le b, \quad c \le y \le d \}  \]
with the sides $AB$, $BC$, $CD$, $DA$, where 
the points $A$, $B$, $C$, $D$ have coordinates
\[ A(a,c), \quad B(a,d), \quad C(b,d), \quad D(b,c) . \]
These sides form a boundary $\partial W$ of $W$. 
Then 
\[
\int_{\partial W} \left( F_x dx +F_y dy \right)=
\int_{BC} F_x dx +\int_{DA} F_x dx +
\int_{AB} F_y dy +\int_{CD} F_y dy =
\]
\[
=\int^b_a F_x(x,d) dx +\int^a_b F_x(x,c) dx +
\int^d_c F_y(a,y) dy +\int^c_d F_y(b,y) dy =
\]
\be \label{Pr1}
=\int^b_a \, dx \, [F_x(x,d) - F_x(x,c)] +
\int^d_c \, dy \, [F_y(a,y) - F_y(b,y) ] .
\ee
The main step of proof of Green's formula is to use 
the Newton-Leibniz formula
\be \label{NLF}
\int^b_a \, dx \, D_x f(x)=f(b)-f(a) .
\ee
The function $f(x)$ in (\ref{NLF}) is absolutely continuous on $[a,b]$. 
As a result, expression (\ref{Pr1}) can be presented as
\[
\int^b_a \, dx \, \left[ \int^d_c \, dy \, D_y F_x(x,y) \right] +
\int^d_c \, dy \, \left[ -\int^b_a \, dx \, D_x F_y(x,y) \right] =
\]
\[
=\int^b_a \, dx \, \int^d_c \, dy \, \left[D_y F_x(x,y)-
D_x F_y(x,y) \right] = \int\int_W dxdy 
\left[D_y F_x -D_x F_y \right] .
\]

To derive a fractional generalization of Green's formula (\ref{Green}), 
we should have a generalization of the Newton-Leibniz formula (\ref{NLF})
in the form
\be \label{20}
_aI^{\alpha}_b \, _aD^{\alpha}_x f(x)=f(b)-f(a) ,
\ee
where some integral and derivative of noninteger order are used.
This generalization exists for specified fractional integrals and derivatives, 
and does not exist for arbitrary taken type of the fractional derivatives.

For the left Riemann-Liouville fractional integral and 
derivative (Lemma 2.5. of \cite{KST}), we have
\be \label{21}
_aI^{\alpha}_b  \, _aD^{\alpha}_x f (x) =f(b)- 
\sum^{n}_{j=1} \frac{(b-a)^{\alpha-j}}{\Gamma(\alpha-j+1)} 
(D^{n-j}_x \ _aI^{n-\alpha}_x f)(a) ,
\ee
where $D^{n-j}_x=d^{n-j}/dx^{n-j}$ are integer derivatives, and $n-1<\alpha<n$.
In particular, if $0<\alpha<1$, then
\be \label{22}
_aI^{\alpha}_b  \, _aD^{\alpha}_x f(x) =f(b)- 
\frac{(b-a)^{\alpha-1}}{\Gamma(\alpha)} \ _aI^{1-\alpha}_b f(x) ,
\ee
Obviously that Eq. (\ref{22}) cannot be considered as a realization of (\ref{20}).
The left Riemann-Liouville fractional integral for $x\in [a,b]$ is defined by
\be
_aI^{\alpha}_x f(x):=\frac{1}{\Gamma (\alpha)} \int^x_a \frac{dx'}{(x-x')^{1-\alpha}} 
\quad (\alpha>0) .
\ee
The left Riemann-Liouville fractional derivative 
for $x\in [a,b]$ and $n-1<\alpha<n$ is defined by
\be
_aD^{\alpha}_x f(x):= D^n_x \, _aI^{n-\alpha}_x f(x) =
\frac{1}{\Gamma (n-\alpha)} \frac{\partial^n}{\partial x^n} \
\int^x_a \frac{f(x') dx'}{(x-x')^{\alpha-n+1}} .
\ee
Note that Eq. (\ref{21}) is satisfied if $f(x)$ is 
Lebesgue measurable functions on $[a,b]$ for which
\[ \int^b_a f(x) \, dx < \infty , \]
and $_aI^{n-\alpha}_b  f(x)$ of the right-hand side of  (\ref{21}) 
has absolutely continuous derivatives up to order $(n-1)$ on $[a,b]$. 


Properties (\ref{21}) and (\ref{22}) 
are connected with the definition of the Riemann-Liouville 
fractional derivative, where the integer-order derivative acts on the fractional integral:
\be \label{RL}
_aD^{\alpha}_x =D^n_x \ _aI^{n-\alpha}_x  , \quad (n-1<\alpha<n) .
\ee
This definition gives that the left--hand side of (\ref{22}) is
\be \label{C1a}
_aI^{\alpha}_x  \ _aD^{\alpha}_x =
\ _aI^{\alpha}_x \, D^n_x \ _aI^{n-\alpha}_x  ,
\ee
where the integer derivative $D^n_x$ is located between the fractional integrals. 
Since the operations $D^n_x$ and $\ _aI^{\alpha}_x$ are not commutative
\[ _aI^{n-\alpha}_x D^n_x -  D^n_x \ _aI^{n-\alpha}_x \ne 0 , \]
we get the additional terms, which cannot give the right-hand side of (\ref{20}).
This noncommutativity can be presented as a nonequivalence of
Riemann-Liouville and Caputo derivatives \cite{Podlubny,KST},
\be
 _a^CD^{\alpha}_x f(x)= \ _aD^{\alpha}_x f(x)-
\sum^{n-1}_{j=0} \frac{(x-a)^{j-\alpha}}{\Gamma(j-\alpha+1)} (D^j_x f)(a) ,
\quad (n-1<\alpha<n) .
\ee
The left Caputo fractional derivative is defined 
by the equation (compare with (\ref{RL}))
\be
_a^CD^{\alpha}_x f(x):= \ _aI^{n-\alpha}_x  D^n_x f(x), 
\quad (n-1<\alpha<n) .
\ee
The noncommutativity of $D^n_x$ and $_aI^{\alpha}_x$ in (\ref{C1a}) does
not allow us to use semi-group property 
(see Lemma 2.3 of \cite{KST} and Theorem 2.5 of \cite{SKM}) 
of fractional integrals 
\be \label{SG}
_aI^{\alpha}_x \ _aI^{\beta}_x = \ _aI^{\alpha+\beta}_x , 
\quad (\alpha>0, \ \beta>0) . \ee
Note that equation (\ref{SG}) is satisfied at almost every point $x\in [a,b]$ for 
$f(x) \in L_p(a,b)$ and $\alpha,\beta>0$.
We denote by $L_p(a,b)$ ($1<p<\infty$) the set of 
those Lebesgue measurable functions on $[a,b]$ for which
\[ \left( \int^b_a dx \, |f(x)|^p \right)^{1/p} < \infty . \]
In general, the semi-group property
\be \label{SGd}
_aD^{\alpha}_x \ _aD^{\beta}_x = \ _aD^{\alpha+\beta}_x , 
\quad (\alpha>0, \ \beta>0) .
\ee
is not satisfied for fractional derivatives 
(see Property 2.4 in \cite{KST}). 
For some special cases, Eq. (\ref{SGd}) can be used 
(see Theorem 2.5. in \cite{SKM}). 
For example, the property (\ref{SGd}) is satisfied for the functions
\[ f(x) \in \, _aI^{\alpha+\beta}_x(L_1(a,b)) , \]
i.e., equation (\ref{SGd}) is valid for $f(x)$
if there exists a function $g(x) \in L_1(a,b)$ such that
\[ f(x)= \, _aI^{\alpha+\beta}_x g(x) . \]
The semi-group property for fractional derivatives is also valid if $a=0$, $b=\infty$ and 
$f(x)$ is infinitely differentiable (generalized) function on $[0,\infty)$
(see Sec.1.4.5. of \cite{GS} and Sec.8.3. of \cite{SKM}).


In order to have a fractional generalization of the Newton-Leibniz formula 
of the form (\ref{20}), we must replace 
the left Riemann-Liouville derivative $\, _aD^{\alpha}_b$
in Eq. (\ref{20}), where  
\[ _aI^{\alpha}_x \ _aD^{\alpha}_x =\, _aI^{\alpha}_b ( D^n_x \ _aI^{n-\alpha}_x )  \]
by the left Caputo derivative $\ _a^CD^{\alpha}_x$, 
such that the left-hand side of (\ref{20}) is
\[ _aI^{\alpha}_x \ _a^CD^{\alpha}_x =
\ _aI^{\alpha}_x ( _aI^{n-\alpha}_x D^n_x ) . \]
Then, we can use the semi-group property (\ref{SG}), and
\[ _aI^{\alpha}_x \ _a^CD^{\alpha}_x f(x)=
\ _aI^{\alpha}_x \ _aI^{n-\alpha}_x D^n_x f(x) =\, _aI^n_x D^n_x f(x) . 
\]
In particular, if $n=1$ and $0<\alpha<1$, then
\be
 _aI^{\alpha}_b \ _a^CD^{\alpha}_x f(x)= \ _aI^1_b \, D^1_x f(x) = 
\int^b_a dx \, D^1_{x} f(x) =f(b)-f(a) .
\ee

As a result, to generalize Gauss's, Green's and Stokes' formulas for fractional case,
we can use the equation with the Riemann-Liouville integral and the Caputo derivative:
\be
_aI^{\alpha}_b \ _a^CD^{\alpha}_x f(x)= f(b)-f(a) .
\ee
This equation can be considered as a fractional analog of the Newton-Leibniz formula.

\section{Fractional Generalization of the Fundamental Theorem of Calculus}

The fundamental theorem of calculus (FTC) is the statement that the 
two central operations of calculus, differentiation and integration, 
are inverse operations: 
if a continuous function is first integrated and then differentiated, 
the original function is retrieved
\be \label{FT1} D^1_x \ _aI^1_x f(x)=f(x) . \ee
An important consequence, sometimes called the second fundamental theorem of calculus, 
allows one to compute integrals by using an antiderivative of 
the function to be integrated:
\be \label{FT2} _aI^1_b \, D^1_x f(x)=f(b)-f(a) .  \ee
If we use the Riemann-Liouville integrals and derivatives \cite{SKM,KST},
we cannot generalize (\ref{FT2}) for fractional case, since
\be 
_aI^{\alpha}_b \ _aD^{\alpha}_x f(x) \ne f(b)-f(a) , 
\ee
In this case, we have equation (\ref{21}).

The FTC states that the integral of a function  $f$ over the interval $[a,b]$ 
can be calculated by finding an antiderivative $F$, i.e., 
a function, whose derivative is $f$. 
Integral theorems of vector calculus (Stokes', Green's, Gauss's theorems) 
can be considered as generalizations of FTC. 

The fractional generalization of the FTC for finite interval $[a,b]$
can be realized (see remarks after proof of the theorem and Section 3.) 
in the following special form. \\

{\bf Fundamental Theorem of Fractional Calculus} \\
{\it (1) Let $f(x)$ be a real-valued function defined on a closed interval $[a, b]$. 
Let $F(x)$ be the function defined for $x$ in $[a, b]$ by
\be \label{T1}
F(x)=\,  _aI^{\alpha}_x  f(x) ,
\ee
where $\ _aI^{\alpha}_x$ is the fractional Riemann-Liouville integral 
\be
 _aI^{\alpha}_x  f(x) :=  \frac{1}{\Gamma(\alpha)} 
\int^{x}_{a} \frac{f(x')}{(x-x')^{1-\alpha}} dx' ,
\ee
then
\be
 _a^CD^{\alpha}_x  F(x) =f(x)
\ee
for $x \in (a, b)$, where $ _a^CD^{\alpha}_x$ is the Caputo fractional derivative
\be
 _a^CD^{\alpha}_x F(x) =\, _aI^{n-\alpha}_x D^n_x F(x)=
\frac{1}{\Gamma(n-\alpha)} \int^{x}_{a} 
\frac{dx'}{(x-x')^{1+\alpha-n}} \frac{d^n F(xý')}{d(x')^n} , 
\quad (n-1<\alpha <n).
\ee

(2)  Let $f(x)$ be a real-valued function defined on a closed interval $[a, b]$. 
Let $F(x)$ be a function such that
\be
f(x) =\, _a^CD^{\alpha}_x  F(x)
\ee
for all $x$ in $[a, b]$, 
then
\be
 _aI^{\alpha}_b  f(x)=F(b)-F(a) ,
\ee
or, equivalently, }
\be \label{FTFC}
 _aI^{\alpha}_b \ _a^CD^{\alpha}_x F(x)=F(b)-F(a) , \quad (0<\alpha<1) .
\ee

\vskip 3 mm

As a result, we have the fractional analogs of 
equations (\ref{FT1}) and (\ref{FT2}) in the form
\be \label{FNLF0}
_a^CD^{\alpha}_x \ _aI^{\alpha}_x f(x)=f(x) , \quad (\alpha>0),
\ee
\be \label{FNLF}
_aI^{\alpha}_x \ _a^CD^{\alpha}_x F(x) = F(x)-F(a) , \quad (0<\alpha<1) , \ee
where $ _aI^{\alpha}_x$ is the Riemann-Liouville integral, and $_a^CD^{\alpha}_x$ is 
the Caputo derivative. \\

{\bf Proof}. \\
The proof of this theorem can be realized by using 
the Lemma 2.21 and Lemma 2.22 of \cite{KST}.

(1) For real values of $\alpha>0$, 
the Caputo fractional derivative 
provides operation inverse to the Riemann-Liouville 
integration from the left (see Lemma 2.21 \cite{KST}), 
\be
 _a^CD^{\alpha}_x \ _aI^{\alpha}_x  f(x) = f(x), \quad (\alpha>0) 
\ee
for $f(x) \in L_{\infty}(a,b)$ or $f(x) \in C[a,b]$.

(2) If $f(x) \in AC^n[a,b]$ or $f(x)\in C^n[a,b]$, then (see Lemma 2.22 \cite{KST})
\be \label{48}
_aI^{\alpha}_x \ _a^CD^{\alpha}_x f(x)=
f(x)-\sum^{n-1}_{j=0} \frac{1}{j!} (x-a)^j (D^j_xf)(a) , \quad (n-1 < \alpha \le n) ,
\ee
where $C^n[a,b]$ is a space of functions, which are $n$ times continuously 
differentiable on $[a,b]$. 
In particular, if $0<\alpha\le 1$ and $f(x) \in AC[a,b]$ or $f(x) \in C[a,b]$, then
\be 
_aI^{\alpha}_x \ _a^CD^{\alpha}_x f(x)= f(x)-f(a).
\ee
This equation can be considered as a fractional generalization 
of the Newton-Leibniz formula in the form (\ref{20}). \\

{\bf Remark 1}. 
In this theorem (see Eqs. (\ref{T1}-\ref{FTFC})), 
the spaces $L_1[a,b]$ and $AC[a,b]$ are used. \\
(a) Here $AC[a,b]$ is a space of functions $F(x)$, 
which are absolutely continuous on $[a,b]$.
It is known that $AC[a,b]$ coincides with the space of primitives of Lebesgue summable
functions and therefore an absolutely continuous function $F(x)$ 
has a summable derivative $D^1_x (x)$ almost everywhere on $[a,b]$.
If $F(x) \in AC[a,b]$, then the Caputo derivative $(0<\alpha<1)$
exists almost everywhere on $[a,b]$ (see Theorem 2.1 of \cite{KST}). \\
(b) We denote $L_p(a,b)$ the set of those Lebesgue measurable 
functions $f$ on $[a,b]$ for which 
\be
\|f\|_p =\left( \int^b_a |f(x)|^p dx  \right)^{1/p} <\infty .
\ee
If $f(x) \in L_p(a,b)$, where $p>1$, then the fractional Riemann-Liouville 
integrations are bounded in $L_p(a,b)$, and 
the semi-group property
\be \label{SG2}
_aI^{\alpha}_x \ _aI^{\beta}_x f(x)= 
\ _aI^{\alpha+\beta}_x f(x) , \quad (\alpha>0, \ \beta>0) 
\ee
are satisfied at almost every point $x \in [a,b]$.
If $\alpha+\beta>1$, then  relations (\ref{SG2}) holds at any point of $[a,b]$
(see Lemma 2.1 and Lemma 2.3 in \cite{KST}). \\

{\bf Remark 2}. 
For the Riemann-Liouville derivative $ _aD^{\alpha}_x$, the relation
\be
 _aD^{\alpha}_x \ _aI^{\alpha}_x  f(x) = f(x), \quad (\alpha>0) 
\ee
holds almost everywhere on $[a,b]$ for $f(x) \in L_p(a,b)$ 
(see Lemma 2.4 of \cite{KST}).

{\bf Remark 3}. 
The Fundamental Theorem of Fractional Calculus (FTFC) 
uses the Riemann-Liouville integration and the Caputo differentiation.
The main property is that the Caputo fractional derivative  
provides us an operation inverse to the Riemann-Liouville 
fractional integration from the left. 
It should be noted that consistent fractional generalizations of the FTC, 
the differential vector operations and 
the integral theorems for other fractional integro-differentiation
such as Riesz, Grunvald-Letnikov, Weyl, Nishimoto are open problems. \\

{\bf Remark 4}. 
In the theorem, we use $0<\alpha\le 1$.
As a result, we obtain the fractional Green's, Stokes' and Gauss's theorems 
for $0<\alpha<1$.
Equation (\ref{FNLF0}) is satisfied for $\alpha \in \mathbb{R}_{+}$.
The Newton-Leibniz formula (\ref{FNLF}) holds for $0<\alpha\le 1$.
For $\alpha>1$, we have (\ref{48}).
As a result, to generalize the Green's, Stokes' and Gauss's theorems for
$\alpha \in \mathbb{R}_{+}$, we can use Eq. (\ref{48}) in the form 
\be \label{48c}
f(b)-f(a)=\, _aI^{\alpha}_b \ _a^CD^{\alpha}_x f(x)+ \sum^{n-1}_{j=1} 
\frac{1}{j!} (b-a)^j f^{(j)} (a) , \quad (n-1<\alpha \le n) , \ee
where $f^{(j)}(x)=D^j_xf(x)$.
In particular, if $1<\alpha \le 2$, then $n=2$ and
\be
f(b)-f(a)=\, _aI^{\alpha}_b \ _a^CD^{\alpha}_x f(x)+ (b-a) f^{\prime}(a) . 
\ee

{\bf Remark 5}. 
In the FTFC, we use the left fractional integrals and derivatives. 
The Newton-Leibniz formulas can be presented for the right fractional Riemann-Liouville 
integrals and the right fractional Caputo derivatives in the form
\be
_xI^{\alpha}_b \, _x^CD^{\alpha}_b f(x)=
f(x)-\sum^{n-1}_{j=0} \frac{(-1)^j f^{(j)} (b)}{j!} (b-x)^j .
\ee
In particular, if $0<\alpha\le 1$, then
\be
_xI^{\alpha}_b \, _x^CD^{\alpha}_b f(x)=f(x)-f(b) .
\ee
For $\alpha>0$, $f(x) \in L_{\infty}(a,b)$ or $f(x) \in C[a,b]$, then
\be
_{x}^CD^{\alpha}_b \, _xI^{\alpha}_b f(x)=f(x) .
\ee
As a result, fractional generalization of differential operations and integral theorems
can be defined for the right integrals and derivatives as well as for the left ones.

\section{Definition of fractional vector operations}

\subsection{Fractional operators}

To define fractional vector operations, we introduce
the operators that correspond to the fractional derivatives and integrals.

We define the fractional integral operator
\be
_aI^{\alpha}_x[x']:=\frac{1}{\Gamma(\alpha)} 
\int^x_a  \frac{dx'}{(x-x')^{1-\alpha}} , \quad (\alpha >0) ,
\ee
which acts on a real-valued function $f(x)\in L_1[a,b]$  by
\be
_aI^{\alpha}_x[x']f(x')=\frac{1}{\Gamma(\alpha)} 
\int^x_a  \frac{f(x')dx'}{(x-x')^{1-\alpha}} .
\ee

The Caputo fractional differential operator on $[a,b]$ can be defined by
\be
_a^CD^{\alpha}_x[x']:=\frac{1}{\Gamma(n-\alpha)} 
\int^x_a  \frac{dx'}{(x-x')^{1+\alpha-n}} 
\frac{\partial^n}{\partial {x'}^n} , \quad (n-1<\alpha<n) ,
\ee
such that the Caputo derivatives for $f(x)\in AC^n[a,b]$ is written as 
\be
_a^CD^{\alpha}_x[x']f(x')=\frac{1}{\Gamma(n-\alpha)} 
\int^x_a  \frac{dx'}{(x-x')^{1+\alpha-n}} 
\frac{\partial^n f(x')}{\partial {x'}^n} , \quad (n-1<\alpha<n) .
\ee
It is easy to see that
\[ _a^CD^{\alpha}_x[x']= _aI^{n-\alpha}_x [x'] D^n[x'] , \quad (n-1<\alpha<n) . \]

Using these notations, formulas (\ref{FNLF0}) and (\ref{FNLF}) 
of the FTFC can be presented as
\be \label{FNLF2a}
_a^CD^{\alpha}_x[x'] \ _aI^{\alpha}_{x'}[x''] f(x'')=f(x) , \quad (\alpha>0) , \ee
\be \label{FNLF2}
_aI^{\alpha}_b[x] \ _a^CD^{\alpha}_x[x'] f(x') = f(b)-f(a) , \quad (0<\alpha<1) . \ee
This form is more convenient than (\ref{FNLF0}) and (\ref{FNLF}),
since it allows us to take into account the variables of integration
and the domain of the operators.

\subsection{Definition of fractional differential vector operations}

Let us define the fractional differential operators (grad, div, curl) 
such that fractional generalizations of 
integral theorems (Green's, Stokes', Gauss') can be realized.
We use the Caputo derivatives to defined these operators and we use
the Riemann-Liouville integrals in the generalizations of the integral theorems. 

Let $W$ be a domain of $\mathbb{R}^3$. 
Let $f(x)$ and ${\bf F}(x)$ be real-valued functions that 
have continuous derivatives up to order $(n-1)$ on $W$, 
such that the $(n-1)$ derivatives are absolutely continuous, 
i.e., $f,{\bf F} \in AC^n[W]$. 
We can define a fractional generalization of nabla operator by
\be \label{fnabla}
\nabla^{\alpha}_W=\, ^C{\bf D}^{\alpha}_W = {\bf e}_1 \, ^CD^{\alpha}_W [x]+
{\bf e}_2 \, ^CD^{\alpha}_W [y]+{\bf e}_3 \, ^CD^{\alpha}_W [z] , \quad (n-1<\alpha < n) .
\ee
Here, we use the fractional Caputo derivatives $ ^CD^{\alpha}_{W}[x_m]$
with respect to coordinates $x_m$.
For the parallelepiped 
\[ W:=\{ a \le x \le b, \quad c \le y \le d, \quad g \le z \le h \} , \]
we have
\be
^CD^{\alpha}_W [x]=\, _a^CD^{\alpha}_b [x] , \quad
^CD^{\alpha}_W [y]=\, _c^CD^{\alpha}_d [y], \quad
^CD^{\alpha}_W [z]=\, _g^CD^{\alpha}_h [z] .
\ee
The right-hand sides of these equations 
the Caputo derivatives are used. \\

(1) If $f=f(x,y,z)$ is $(n-1)$ times continuously differentiable 
scalar field such that $D^{n-1}_{x_l} f$ is absolutely continuous, then
we define its fractional gradient as the following
\[
Grad^{\alpha}_W f=\,  ^C{\bf D}^{\alpha}_W f=
 {\bf e}_l \ ^CD^{\alpha}_W[x_l] f(x,y,z)=
\]
\be
= {\bf e}_1 \ ^CD^{\alpha}_W[x] f(x,y,z)+ {\bf e}_2  
\ ^CD^{\alpha}_W[y] f(x,y,z)+  {\bf e}_3 \ ^CD^{\alpha}_W[z] f(x,y,z) .
\ee

(2) If ${\bf F}(x,y,z)$ is $(n-1)$ times continuously differentiable 
vector field such that $D^{n-1}_{x_l} F_l$ are absolutely continuous, 
then we define its fractional divergence as a value of the expression
\[
Div^{\alpha}_W {\bf F}= \Bigl(\, ^C{\bf D}^{\alpha}_W , {\bf F}\Bigr)=
\, ^CD^{\alpha}_W[x_l] F_l(x,y,z)=
\]
\be
=\, ^CD^{\alpha}_W[x] F_x(x,y,z)+
\, ^CD^{\alpha}_W[y] F_y(x,y,z)+ \ ^CD^{\alpha}_W[z] F_z(x,y,z) .
\ee

(3) The fractional curl operator is defined by
\[
Curl^{\alpha}_W {\bf F}=\Bigl[\, ^C{\bf D}^{\alpha}_W, {\bf F} \Bigr]=
{\bf e}_l \varepsilon_{lmk} \, ^CD^{\alpha}_{W} [x_m] F_k= 
{\bf e}_1 \left(\ ^CD^{\alpha}_{W} [y] F_z-\ ^CD^{\alpha}_{W} [z] F_y \right) +
\]
\be
+{\bf e}_2  \left(\ ^CD^{\alpha}_{W} [z] F_x-\ ^CD^{\alpha}_{W} [x] F_z \right) 
+{\bf e}_3  \left(\ ^CD^{\alpha}_{W} [x] F_y-\ ^CD^{\alpha}_{W} [y] F_x \right) ,
\ee
where $F_k=F_k(x,y,z)\in AC^n[W]$, $(k=1,2,3)$. 

Note that these fractional differential operators are nonlocal.
As a result, the fractional gradient, divergence and curl
depend on the region $W$.

\subsection{Relations for fractional differential vetor operations}

(a) The first relation for the scalar field $f=f(x,y,z)$ is
\be
Curl^{\alpha}_W \, Grad^{\alpha}_W f=
{\bf e}_l \, \varepsilon_{lmn} 
\ ^CD^{\alpha}_W [x_m] \, ^CD^{\alpha}_W[x_n] f =0 ,
\ee
where $\varepsilon_{lmn}$ is Levi-Civita symbol, i.e. 
it is $1$ if $ (i, j, k)$ is an even permutation of $(1,2,3)$,  
$(-1)$ if it is an odd permutation, and $0$ if any index is repeated.

(b) The second relation,
\be
Div^{\alpha}_W \, Grad^{\alpha}_W f(x,y,z) =
\ ^CD^{\alpha}_W [x_l] \ ^CD^{\alpha}_W[x_l] f(x,y,z) =
\sum^3_{l=1}( ^CD^{\alpha}_W [x_l])^2 f(x,y,z) .
\ee
Using notation (\ref{fnabla}),
\be
Div^{\alpha}_W \, Grad^{\alpha}_W  =(\, ^C{\bf D}^{\alpha}_W )^2 
=\Bigl(\, ^C{\bf D}^{\alpha}_W,\, ^C{\bf D}^{\alpha}_W \Bigr) .
\ee
In the general case,
\be
( ^CD^{\alpha}_W [x_l])^2 \ne \ ^CD^{2 \alpha}_W [x_l] .
\ee
It is obvious from
\[ (\, _a^CD^{\alpha}_x )^2=\, _aI^{n-\alpha}_x D^n_x \ _aI^{n-\alpha}_x D^n_x = \]
\[ =\ _aI^{n-\alpha}_x \ _aI^{n-\alpha}_x D^n_x D^n_x +
\, _aI^{n-\alpha}_x [D^n_x, \ _aI^{n-\alpha}_x] D^n_x =
\ _aD^{2\alpha}_x + \, _aI^{n-\alpha}_x [D^n_x, \ _aI^{n-\alpha}_x] D^n_x 
, \]
where
\[  [D^n_x, \ _aI^{n-\alpha}_x] :=
D^n_x \, _aI^{n-\alpha}_x - \ _aI^{n-\alpha}_x D^n_x =
\ _aD^{\alpha}_x - \, _a^CD^{\alpha}_x \ne 0. \]

(c) It is easy to prove the following relation,
\[
Div^{\alpha}_W \, Curl^{\alpha}_W {\bf F}(x,y,z) = \ ^CD^{\alpha}_W [x_l] 
\varepsilon_{lmn} \ ^CD^{\alpha}_W[x_m] F_n(x,y,z) = \]
\be
=\varepsilon_{lmn} \ ^CD^{\alpha}_W [x_l] \ ^CD^{\alpha}_W[x_m] F_n(x,y,z) =0 ,
\ee
where we use antisymmetry of $\varepsilon_{lmn}$ with respect to $m$ and $n$.

(d) There exists a relation for the double curl operation in the form
\[
Curl^{\alpha}_W \, Curl^{\alpha}_W {\bf F}(x,y,z) = {\bf e}_l
\varepsilon_{lmn} \ ^CD^{\alpha}_W[x_m] 
\varepsilon_{npq} \ ^CD^{\alpha}_W[x_p] F_q(x,y,z) = 
\]
\be 
= {\bf e}_l \varepsilon_{lmn} \varepsilon_{npq} 
\ ^CD^{\alpha}_W[x_m] \ ^CD^{\alpha}_W[x_p] F_q(x,y,z) .
\ee
Using
\be
\varepsilon_{lmn} \varepsilon_{lpq} =\delta_{mp} \delta_{nq}-\delta_{mq} \delta_{np} ,
\ee
we get
\be \label{CC}
Curl^{\alpha}_W \, Curl^{\alpha}_W \, {\bf F}(x,y,z) = 
Grad^{\alpha}_W \, Div^{\alpha}_W \, {\bf F}(x,y,z) -
(\, ^C{\bf D}^{\alpha}_W)^2  {\bf F}(x,y,z) ,
\ee

(e) In the general case,
\be
_a^CD^{\alpha}_x [x'] \Bigr(f(x') g(x')\Bigr) \ne 
\Bigr( \, _a^CD^{\alpha}_x[x']  f(x') \Bigr) g(x)+
\Bigr(\, _a^CD^{\alpha}_x[x']  g(x') \Bigr) f(x) .
\ee
For example (see Theorem 15.1. from \cite{SKM}),
\be
_aD^{\alpha}_x[x'] \Bigr(f(x') g(x')\Bigr) =
\sum^{\infty}_{j=0} \frac{\Gamma(\alpha+1)}{\Gamma(j+1)\Gamma(\alpha-j+1)} 
\Bigr(\, _aD^{\alpha-j}_x[x'] f(x') \Bigr) \Bigr(\, D^j_x g(x)\Bigr) ,
\ee 
if $f(x)$ and $g(x)$ are analytic functions on $[a,b]$. 
As a result, we have
\be
Grad^{\alpha}_W \Bigl( f g \Bigr) \ne
\Bigl( Grad^{\alpha}_W f \Bigr) g+
\Bigl( Grad^{\alpha}_W g \Bigr) f ,
\ee
\be
Div^{\alpha}_W \Bigl( f {\bf F} \Bigr) \ne
\Bigl( Grad^{\alpha}_W f , {\bf F} \Bigr)+
f \ Div^{\alpha}_W {\bf F} .
\ee
These relations state that we cannot use the Leibniz rule 
in a fractional generalization of the vector calculus.

\subsection{Fractional integral vector operations}

In this section, we define fractional generalizations of circulation, 
flux and volume integral.

Let ${\bf F}={\bf F}(x,y,z)$ be a vector field such that
\[ {\bf F}(x,y,z)={\bf e}_1 F_x(x,y,z)+ {\bf e}_2 F_y(x,y,z)+ {\bf e}_3 F_z(x,y,z) . \]
If $F_x$, $F_y$, $F_x$ are absolutely integrable real-valued functions on $\mathbb{R}^3$, 
i.e.,  $F_x$, $F_y$, $F_x \in L_1(\mathbb{R}^3)$, then we can define 
the following fractional integral vector operations of order $\alpha>0$.

(1) A fractional circulation is a fractional line integral along a line $L$ 
that is defined by
\be
{\cal E}^{\alpha}_L({\bf F})= \Bigl( {\bf I}^{\alpha}_L, {\bf F} \Bigr)=
I^{\alpha}_L [x] F_x+I^{\alpha}_L [y] F_y+I^{\alpha}_L [z] F_z .
\ee
For $\alpha=1$, we get
\be
{\cal E}^1_L({\bf F})= \Bigl( {\bf I}^{1}_L, {\bf F} \Bigr)=
\int_L \Bigl( {\bf dL} , {\bf F} \Bigr)=\int_L (F_x dx+ F_y dy+F_z dz) ,
\ee
where ${\bf dL}= {\bf e}_1 dx+ {\bf e}_2 dy+  {\bf e}_3 dz$. \\

(2) A fractional flux of the vector field ${\bf F}$ across a surface $S$ 
is a fractional surface integral of the field, such that
\be
{\Phi}^{\alpha}_S ({\bf F})= \Bigl( {\bf I}^{\alpha}_S, {\bf F} \Bigr)=
I^{\alpha}_S [y,z] F_x+I^{\alpha}_S [z,x] F_y+I^{\alpha}_S [x,y] F_z .
\ee
For $\alpha=1$, we get
\be
{\Phi}^1_S ({\bf F})=  \Bigl( {\bf I}^1_S, {\bf F} \Bigr)=
\int\int_S \Bigl(d {\bf S}, {\bf F}\Bigr)=
\int \int_S (F_x dydz + F_y dzdx + F_z dxdy) ,
\ee
where ${\bf dS}= {\bf e}_1 dy dz+ {\bf e}_2 dz dx +  {\bf e}_3 dx dy$. \\

(3) A fractional volume integral is a triple fractional 
integral within a region $W$ in $\mathbb{R}^3$ of a scalar field $f= f(x,y,z)$, 
\be
V^{\alpha}_{W}(f) =I^{\alpha}_W[x,y,z] f(x,y,z)=
I^{\alpha}_W [x] I^{\alpha}_W [y] I^{\alpha}_W [z] f(x,y,z) .
\ee
For $\alpha=1$, we have
\be
V^{1}_{W}(f):=\int\int\int_W \, dV \, f(x,y,z)=\int\int\int_W \, dxdydz \, f(x,y,z) .
\ee
This is the usual volume integral for the function $f(x,y,z)$.

\section{Fractional Green's formula}

Green's theorem gives the relationship between a line integral around 
a simple closed curve $\partial W$
and a double integral over the plane region $W$ bounded by $\partial W$. 
The theorem statement is the following. 
Let $\partial W$ be a positively oriented, piecewise smooth, simple closed curve 
in the plane and let $W$ be a region bounded by $\partial W$.  
If $F_x$ and $F_y$ have continuous partial derivatives 
on an open region containing $W$, then
\be \label{GF}
\int_{\partial W} \Bigl( F_x dx +F_y dy \Bigr)=
\int \int_W  \Bigl( D_y F_x- D_x F_y \Bigr) dxdy .
\ee
A fractional generalization of the Green's formula (\ref{GF}) 
is presented by the following statement. \\

{\bf Theorem} (Fractional Green's Theorem for a Rectangle)\\
{\it Let $F_x(x,y)$ and $F_y(x,y)$ be absolutely continuous 
(or continuously differentiable) real-valued functions
in a domain that includes the rectangle 
\be \label{rectangle}
W:=\{(x,y) : \quad a\le x\le b, \quad c \le y \le d \}  .
\ee
Let the boundary of $W$ be the closed curve $\partial W$. Then
\be \label{FGreen}
I^{\alpha}_{\partial W} [x] F_x(x,y)+ I^{\alpha}_{\partial W} [y] F_y(x,y) =
\ I^{\alpha}_{W} [x,y] 
\left(\, ^CD^{\alpha}_{\partial W} [y] F_x(x,y)-\, ^CD^{\alpha}_{\partial W} [x] F_y(x,y) \right) , 
\ee
where $0<\alpha\le 1$. } \\

{\bf Proof} \\
To prove equation (\ref{FGreen}), we change the double fractional integral 
$I^{\alpha}_{W} [x,y]$ to the repeated fractional integrals
$I^{\alpha}_W [x] \ I^{\alpha}_W [y]$,
and then employ the Fundamental Theorem of Fractional Calculus. 

Let $W$ be the rectangular domain (\ref{rectangle})
with the sides $AB$, $BC$, $CD$, $DA$, where 
the points $A$, $B$, $C$, $D$ have coordinates
\[ A(a,c), \quad B(a,d), \quad C(b,d), \quad D(b,c) . \]
These sides form the boundary $\partial W$ of $W$. 

For the rectangular region $W$ defined by $a\le x\le b$, $c \le y \le d $,
the repeated integral is
\[ I^{\alpha}_W [x] \ I^{\alpha}_W [y]=\, _aI^{\alpha}_b [x] \ _cI^{\alpha}_d [y] , \]
and equation (\ref{FGreen}) is
\[
_aI^{\alpha}_b [x] \left( F_x(x,d)-F_x(x,c) \right) + \,
_cI^{\alpha}_d [y] \left( F_y(a,y)-F_y(b,y) \right) =
\]
\be \label{93}
= \ _aI^{\alpha}_b [x] \ _cI^{\alpha}_d [y] 
\left( \ _c^CD^{\alpha}_y [y'] F_x(x,y') - \ _a^CD^{\alpha}_x [x'] F_y(x',y) \right).
\ee
To prove of the fractional Green's formula, 
we realize the following transformations
\[
\Bigl( {\bf I}^{\alpha}_{\partial W} ,{\bf F} \Bigr)
= I^{\alpha}_{\partial W}[x] F_x+I^{\alpha}_{\partial W}[y] F_y=
I^{\alpha}_{BC}[x] F_x + I^{\alpha}_{DA} [x] F_x +
I^{\alpha}_{AB}[y] F_y + I^{\alpha}_{CD} [y] F_y =
\]
\[
=\, _aI^{\alpha}_b [x] F_x(x,d)  -\, _aI^{\alpha}_b [x] F_x(x,c) + 
\, _cI^{\alpha}_d [y] F_y(a,y) dy - \, _cI^{\alpha}_d [y] F_y(b,y) =
\]
\be \label{Pr2}
=\, _aI^{\alpha}_b[x] \, [F_x(x,d) - F_x(x,c)] +
\, _cI^{\alpha}_d [y] \, [F_y(a,y) - F_y(b,y)] .
\ee
The main step of the proof of Green's formula is to use the
fractional Newton-Leibniz formula
\[
F_x(x,d)-F_x(x,c)=\, _cI^{\alpha}_d [y]\, _c^CD^{\alpha}_y[y'] \, F(x,y') , \]
\be
F_y(a,y)-F_y(b,y)= -\,  _aI^{\alpha}_b[x] \, _a^CD^{\alpha}_x[x'] \, F(x',y) .
\ee
As a result, expression (\ref{Pr2}) can be presented as
\[
_aI^{\alpha}_b [x] \, \Bigl\{ \, _cI^{\alpha}_d [y]\, _c^CD^{\alpha}_y [y'] \, F_x(x,y') \Bigr\} +
_cI^{\alpha}_d [y] \, \Bigl\{ - \, _aI^{\alpha}_b [x] \, _a^CD^{\alpha}_x [x'] \, F_y(x',y) \Bigr\} =
\]
\[
=\, _aI^{\alpha}_b [x] \,  _cI^{\alpha}_d [y] \, \Bigl(\, _c^CD^{\alpha}_y [y'] \, F_x(x,y') -
\, _a^CD^{\alpha}_x [x'] \, F_y(x',y) \Bigr) =
\]
\[
=I^{\alpha}_W [x,y] \,   \Bigl(\, _c^CD^{\alpha}_y [y'] \, F_x(x,y') -
\, _a^CD^{\alpha}_x [x'] \, F_y(x',y) \Bigr) .
\]
This is the left-hand side of Eq. (\ref{93}).
This ends the proof. \\

{\bf Remark 1}. 
In this fractional Green's theorem, we use the rectangular region $W$. 
If the region can be approximated by a set of rectangles, 
the fractional Green's formula can also be proved.
In this case, the boundary $\partial W$ is presented by paths each consisting of 
horizontal and vertical line segments, lying in $W$. \\

{\bf Remark 2}. 
To define the double integral and the theorem for nonrectangular regions $R$,
we can consider the function ${\bf f}(x,y)$, that is defined in 
the rectangular region $W$ such that $R \subset W$ and
\be
{\bf f}(x,y)=
\begin{cases} 
{\bf F}(x,y) , & (x,y) \in R ;
\cr 
0, & (x,y) \in W/R  .
\end{cases}
\ee
As a result, we define
a fractional double integral over the nonrectangular region $R$,
through the fractional double integral over the rectangular region $W$:
\be
{\bf I}^{\alpha}_{R} [x,y] \, {\bf F}(x,y) = {\bf I}^{\alpha}_{W} [x,y] \, {\bf f}(x,y) . 
\ee

To define double integrals over nonrectangular regions,
we can use a fairly general method to calculate them.
For example, we can do this for special regions called elementary regions.
Let $R$ be a set of all points $(x,y)$ such that
\[ a \le x \le b , \quad \varphi_1 (x) \le y \le \varphi_2 (x) . \]
Then, the double integrals for such regions can be calculated by
\be
{\bf I}^{\alpha}_{R} [x,y] \, F(x,y) =
\, _a{\bf I}^{\alpha}_b [x] 
\ _{\varphi_1(x)}{\bf I}^{\alpha}_{\varphi_2 (x)}[y] \, F(x,y) .
\ee
It is easy to consider the following examples. \\
1) $ \varphi_1 (x)=0$, $y=\varphi_2 (x)=x^2$, $F(x,y)=x+y$. \\
2) $ \varphi_1 (x)=x^3$, $ \varphi_1 (x)=x^2$,  $F(x,y)=x+y$. \\
3) $ \varphi_1 (x)=0$, $y=\varphi_2 (x)=x$, $F(x,y)=xy$. \\
The fractional integrals can be calculated by using the relations 
\be
_aI^{\alpha}_x [x] (x-a)^{\beta} =
\frac{\Gamma(\beta+1)}{\Gamma(\beta+\alpha+1)} (x-a)^{\beta+\alpha} ,
\ee
where $\alpha>0$, $\beta>0$.
For other relations see Table 9.1 in \cite{SKM}.
To calculate the Caputo derivatives, we can use this table and the equation
\be
_a^CD^{\alpha}_x [x'] f(x') =\, _aD^{\alpha}_x [x'] f(x')-
\sum^{n-1}_{k=0} \frac{f^{(k)}(a)}{\Gamma(k-\alpha+1)}, 
\quad n-1< \alpha \le n .
\ee
Note that the Mittag-Leffler function $E_{\alpha}[ (x'-a)^{\alpha}]$
is not changed by the Caputo derivative 
\be
_a^CD^{\alpha}_x [x'] \, E_{\alpha} [ (x'-a)^{\alpha}]= 
E_{\alpha} [ (x-a)^{\alpha}].
\ee
This equation is a fractional analog of the well-known property 
of exponential function of the form $D^1_x \exp(x-a) =\exp(x-a)$. 
Therefore the Mittag-Leffler function can be considered as 
a fractional analog of exponential function.

\section{Fractional Stokes' formula}


We shall restrict ourselves to the consideration of a simple surface.
If we denote the boundary of the simple surface $W$ by $\partial W$
and if ${\bf F}$ is a smooth vector field defined on $W$,
then the Stokes' theorem asserts that
\be \label{94}
\int_{\partial W} \Bigl( {\bf F}, d{\bf L} \Bigr)=
\int_{W} \Bigl( curl {\bf F}, d {\bf S} \Bigr) .
\ee
The right-hand side of this equation is the surface integral of
$curl {\bf F}$ over $W$, whereas the left-hand side is 
the line integral of ${\bf F}$ over the line $\partial W$.
Thus the  Stokes' theorem is the assertion that the line integral of a vector field
over the boundary of the surface $W$ is the same as the integral over the surface
of the curl of ${\bf F}$. 

For Cartesian coordinates, Eq. (\ref{94}) gives
\[
\int_{\partial W} \Bigl( F_x dx+F_y dy+ F_z dz \Bigr)=
\]
\be \label{St}
=\int \int_{W} \ \Bigl( dy dz \, [D_y F_z-D_z F_y] +
 dz dx \, [D_z F_x-D_x F_z] + 
\ dx dy \, [D_x F_y-D_y F_x] \Bigr) .
\ee
Let ${\bf F}={\bf F}(x,y,z)$ be a vector field such that
\[ {\bf F}(x,y,z)={\bf e}_1 F_x(x,y,z)+ {\bf e}_2 F_y(x,y,z)+ {\bf e}_3 F_z(x,y,z) . \]
where $F_x$, $F_y$, $F_x$ are 
absolutely continuous (or continuously differentiable) 
real-valued functions on $\mathbb{R}^3$.
Then the fractional generalization of the Stokes' formula (\ref{St}) can be written as
\be \label{FS}
\Bigl( {\bf I}^{\alpha}_{\partial W} , {\bf F} \Bigr) =
\Bigl( {\bf I}^{\alpha}_{W} , Curl^{\alpha}_{\partial W} {\bf F} \Bigr) .
\ee
Here we use the notations
\be \label{Int1}
{\bf I}^{\alpha}_{L}=
{\bf I}^{\alpha}_{\partial W} ={\bf e}_m I^{\alpha}_{\partial W}[x_m]=
{\bf e}_1 I^{\alpha}_{\partial W}[x]+
{\bf e}_2 I^{\alpha}_{\partial W}[y]+
{\bf e}_3 I^{\alpha}_{\partial W}[z] ,
\ee
such that
\be
\Bigl( {\bf I}^{\alpha}_{\partial W} , {\bf F} \Bigr) =
I^{\alpha}_{\partial W} [x] F_x + 
I^{\alpha}_{\partial W} [y] F_y +
I^{\alpha}_{\partial W} [z] F_z .
\ee
The integral (\ref{Int1}) can be considered as a fractional line integral.

In the right-hand side of (\ref{FS}), ${\bf I}^{\alpha}_W$ is
a fractional surface integral over $S=W$ such that
\be
{\bf I}^{\alpha}_{S} =
{\bf I}^{\alpha}_{W} = {\bf e}_1 I^{\alpha}_{W} [y,z] + {\bf e}_2 I^{\alpha}_{W} [z,x] 
+{\bf e}_3 I^{\alpha}_{W} [x,y] .
\ee
The fractional curl operation is
\[
Curl^{\alpha}_W {\bf F}={\bf e}_l \varepsilon_{lmn} \, ^CD^{\alpha}_{W} [x_m] F_n= 
{\bf e}_1 \left(\ ^CD^{\alpha}_{W} [y] F_z-\ ^CD^{\alpha}_{W} [z] F_y \right) +
\]
\be \label{curl-a}
+{\bf e}_2  \left(\ ^CD^{\alpha}_{W} [z] F_x-\ ^CD^{\alpha}_{W} [x] F_z \right) 
+{\bf e}_3  \left(\ ^CD^{\alpha}_{W} [x] F_y-\ ^CD^{\alpha}_{W} [y] F_x \right) .
\ee
For $\alpha =1$, equation (\ref{curl-a}) gives the well-known expression
\[ Curl^{1}_W {\bf F}=curl {\bf F}=
{\bf e}_l \varepsilon_{lmn} \, D_{x_m} F_n= 
{\bf e}_1 \left( D_y F_z-D_z F_y \right) + \]
\be
+{\bf e}_2  \left( D_z F_x- D_x F_z \right) 
+{\bf e}_3  \left( D_x F_y- D_y F_x \right) .
\ee
The right-hand side of Eq. (\ref{FS}) means
\[
\Bigl( {\bf I}^{\alpha}_{W} , Curl^{\alpha}_{W} {\bf F} \Bigr)=
I^{\alpha}_{W} [y,z] \left(\ ^CD^{\alpha}_{W} [y] F_z-\ ^CD^{\alpha}_{W} [z] F_y \right) +
\]
\be
+I^{\alpha}_{W} [z,x] \left(\ ^CD^{\alpha}_{W} [z] F_x-
\ ^CD^{\alpha}_{W} [x] F_z \right) +
I^{\alpha}_{W} [x,y] \left(\ ^CD^{\alpha}_{W} [x] F_y-
\ ^CD^{\alpha}_{W} [y] F_x \right) .
\ee
This integral can be considered as a fractional surface integral.

\section{Fractional Gauss's formula}

Let us give the basic theorem
regarding the Gauss's  formula in a fractional case. \\

{\bf Theorem} (Fractional Gauss's Theorem for a Parallelepiped)\\
{\it Let $F_x(x,y,z)$, $F_y(x,y)$, $F_z(x,y,z)$ 
be continuously differentiable real-valued functions
in a domain that includes the parallelepiped
\be \label{rectangle2}
W :=\{(x,y,z) : \quad a\le x\le b, \quad c\le y \le d , \quad g \le z \le h \} .
\ee
If the boundary of $W$ be a closed surface $\partial W$, then
\be
\Bigl( {\bf I}^{\alpha}_{\partial W} , {\bf F} \Bigr)= 
I^{\alpha}_W Div^{\alpha}_W {\bf F} .
\ee
This equation can be called the fractional Gauss's formula.} \\

{\bf Proof}. \\
For Cartesian coordinates, we have the vector field
${\bf F}= F_x {\bf e}_1 + F_y {\bf e}_2 + F_z {\bf e}_3$, and
\be \label{67}
I^{\alpha}_{W}= I^{\alpha}_{W}[x,y,z], \quad
{\bf I}^{\alpha}_{\partial W} =
{\bf e}_1 I^{\alpha}_{\partial W}[y,z]+ 
{\bf e}_2 I^{\alpha}_{\partial W}[x,z]+
{\bf e}_3 I^{\alpha}_{\partial W}[x,y] .
\ee
Then
\be
\Bigl( {\bf I}^{\alpha}_{\partial W} , {\bf F} \Bigr)= 
I^{\alpha}_{\partial W}[y,z] F_x+
I^{\alpha}_{\partial W}[x,z] F_y+
I^{\alpha}_{\partial W}[x,y] F_z  ,
\ee
and
\be
I^{\alpha}_W Div^{\alpha}_W {\bf F}=
I^{\alpha}_{W}[x,y,z] \Bigl( \, ^CD^{\alpha}_{\partial W}[x] F_x+
\, ^CD^{\alpha}_{\partial W}[y] F_y+ \, ^CD^{\alpha}_{\partial W}[z] F_z \Bigr) .
\ee
If $W$ is the parallelepiped 
\be
W:=\{ a\le x\le b, \quad c\le y \le d , \quad g \le z \le h \} ,
\ee
then the integrals (\ref{67}) are
\be
I^{\alpha}_{W}[x,y,z]=\, _aI^{\alpha}_b [x] \ _cI^{\alpha}_d [y] \ _gI^{\alpha}_h [z] ,
\ee
and
\be
I^{\alpha}_{\partial W}[y,z]=\, _cI^{\alpha}_d[y] \ _gI^{\alpha}_h [z] ,
\ee
\be
I^{\alpha}_{\partial W}[x,z]=\, _aI^{\alpha}_b[x] \ _gI^{\alpha}_h [z] ,
\ee
\be
I^{\alpha}_{\partial W}[x,y]=\, _aI^{\alpha}_b[x] \ _cI^{\alpha}_d [y] .
\ee
As a result, we can realize the following transformations
\[
\Bigl( {\bf I}^{\alpha}_{\partial W}, {\bf F} \Bigr) =
I^{\alpha}_{\partial W}[y,z] F_x + 
I^{\alpha}_{\partial W}[z,x] F_y + 
I^{\alpha}_{\partial W}[x,y] F_z = 
\]
\[
=\, _cI^{\alpha}_d [y]\, _gI^{\alpha}_h [z] \, \Bigl\{ F_x(b,y,z)-F_x(a,y,z) \Bigr\}+\,
_aI^{\alpha}_b [x] \, _gI^{\alpha}_h [z] \, \Bigl\{F_y(x,d,z)-F_y(x,c,z) \Bigr\}+ \]
\[ +\, _aI^{\alpha}_b [x] _cI^{\alpha}_d [y] \, \Bigl\{F_z(x,y,g)-F_z(x,y,h)\Bigr\}= \]
\[
=\, _aI^{\alpha}_b [x] \, _cI^{\alpha}_d [y] \, _gI^{\alpha}_h [z]
\, \Bigl\{ \, _a^CD^{\alpha}_x [x']  F_x(x',y,z)+ \, _c^CD^{\alpha}_y [y'] F_y(x,y',z)+ 
\, _g^CD^{\alpha}_z [z'] F_z(x,y,z') \Bigr\} =
\]
\[
= I^{\alpha}_W  \Bigr(\, ^C{\bf D}^{\alpha}_W , {\bf F} \Bigr) =
I^{\alpha}_W Div^{\alpha}_W {\bf F} .
\]
This ends the proof of 
the fractional Gauss's formula for parallelepiped region. \\

{\bf Remark}. 
To define the triple integral and the theorem for non-parallelepiped regions $R$,
we consider the function $f(x,y,z)$, that is defined in 
the parallelepiped region $W$ such that $R \subset W$, such that
\be
f(x,y,z)=
\begin{cases} 
F(x,y,z) , & (x,y,z) \in R ;
\cr 
0, & (x,y,z) \in W/R  .
\end{cases}
\ee
Then we have
\be
{\bf I}^{\alpha}_{R} [x,y,z] F(x,y,z) = {\bf I}^{\alpha}_{W} [x,y,z] f(x,y,z) . 
\ee
As a result, we define
a fractional triple integral over the non-parallelepiped region $R$,
through the fractional triple integral over the parallelepiped region $W$.

\section{Fractional differential forms}

\subsection{Brief description of different approaches}

A fractional generalization of differential has 
been presented by Ben Adda in \cite{BA1,BA2}.
A fractional generalization of the differential forms has been 
suggested by Cottrill-Shepherd and Naber in \cite{FDF} (see also \cite{YZH,Kaz}).
The application of fractional differential forms to
dynamical systems are considered in \cite{T2,T3}.
Fractional integral theorems are not considered.

(1) In the papers \cite{BA1,BA2}, the fractional differential 
for analytic functions is defined by
\be
d^{\alpha} f=\frac{1}{\Gamma(1+\alpha)} dx_j N^{\alpha}_{x_j} f(x) ,
\ee
where $N^{\alpha}_{x_j}$ are Nishimoto fractional derivatives \cite{N} 
(see also Section 22 of \cite{SKM}), which is a generalization of 
the Cauchy's differentiation formula.

(2) In the paper \cite{FDF} (see also \cite{YZH,T2,Kaz}), 
an exterior fractional differential 
is defined through the Riemann-Liouville derivatives by
\be
d^{\alpha}=\sum^n_{j=1} (dx_j)^{\alpha} \, _0D^{\alpha}_{x_j} .
\ee
In two dimensions $n=2$, 
\[ d^{\alpha}=(dx)^{\alpha}\, _0D^{\alpha}_x+ (dy)^{\alpha} \, _0D^{\alpha}_y , \]
suhc that
\be \label{dx1}
d^{\alpha}x=(dx)^{\alpha} \frac{x^{1-\alpha}}{\Gamma(2-\alpha)} + 
(dy)^{\alpha} \frac{x y^{-\alpha}}{\Gamma(1-\alpha)} ,
\ee
\be \label{dy1}
d^{\alpha}y=(dx)^{\alpha} \frac{y x^{-\alpha}}{\Gamma(1-\alpha)} + 
(dy)^{\alpha} \frac{y^{1-\alpha}}{\Gamma(2-\alpha)} ,
\ee
where we use
\be
_0D^{\alpha}_{x_j} 1= \frac{ x^{-\alpha}_j}{\Gamma(1-\alpha)} . 
\ee

(3) In the paper \cite{T3}, an exterior fractional differential 
is defined through the fractional Caputo derivatives in the form
\be \label{VT}
d^{\alpha}=\sum^n_{j=1} (dx_j)^{\alpha} \, _{0}^CD^{\alpha}_{x_j} .
\ee
For two dimensions $(x,y)$, we have
\[ d^{\alpha}=(dx)^{\alpha}\, _0^CD^{\alpha}_x+ (dy)^{\alpha} \, _0^CD^{\alpha}_y , \]
such that
\be
d^{\alpha}x=(dx)^{\alpha} \frac{x^{1-\alpha}}{\Gamma(2-\alpha)} ,
\ee
\be
d^{\alpha}y=(dy)^{\alpha} \frac{y^{1-\alpha}}{\Gamma(2-\alpha)} ,
\ee
(compare with (\ref{dx1}) and (\ref{dy1})).
Equation (\ref{VT}) can be rewritten as
\be
d^{\alpha}=\sum^n_{j=1} \Gamma(2-\alpha) x^{\alpha-1}_j d^{\alpha} x_j \,  _0^CD^{\alpha}_{x_j} .
\ee
This relation is used in \cite{T3} as a fractional exterior differential.

\subsection{Definition of a fractional exterior differential}

A definition of fractional differential forms must be correlated 
with a possible generalization of the fractional integration of differential forms.
To derive fractional analogs of differential forms
and its integrals, we consider a simplest case that is 
an exact 1-form on the interval $L=[a,b]$.
It allows us to use the fractional Newton-Leibniz formula.

In order to define an integration of fractional differential forms, 
we can use the fractional Riemann-Liouville integrals.
Then a fractional exterior derivative must be defined 
through the Caputo fractional derivative.

Equation (\ref{FNLF2}) of FTFC means that 
\be \label{119}
\int^x_a \frac{dx'}{\Gamma(\alpha) (x-x')^{1-\alpha}} \ _a^CD^{\alpha}_{x'}[x''] f(x'') 
= f(x)-f(a) , \quad (0<\alpha<1) . 
\ee
Using
\[ dx'=sgn(dx') |dx'|=sgn(dx') |dx'|^{1-\alpha} |dx'|^{\alpha}, \quad (0<\alpha<1) , \]
equation (\ref{119}) can be presented in the form
\be \label{120}
\int^x_a \frac{|dx'|^{1-\alpha}}{\Gamma(\alpha) (x-x')^{1-\alpha}} \ 
\Bigl( sgn(dx') |dx'|^{\alpha} \, _a^CD^{\alpha}_{x'}[x''] f(x'') \Bigr) 
= f(x)-f(a) , \quad (0<\alpha<1) . 
\ee
The expression in the big brackets of (\ref{120}) 
can be considered as a fractional differential of the function $f(x)$. 
As a result, we have
\be \label{Idf}
\hat I^{\alpha}_L [x] \, _ad^{\alpha}_x f(x) =f(b)-f(a) , \quad (0<\alpha<1),
\ee
where $L=[a,b]$, and 
the fractional integration for differential forms is defined by the operator
\be
\hat I^{\alpha}_L[x] := \int^b_a \frac{|dx|^{1-\alpha}}{\Gamma(\alpha) (b-x)^{1-\alpha}} .
\ee
The exact fractional differential 0-form is a fractional differential of the function
\be
_ad^{\alpha}_x f(x):= (dx)^{\alpha} \, _a^CD^{\alpha}_{x}[x'] f(x') .
\ee
Equation (\ref{Idf}) can be considered as a fractional generalization
of the integral for differential 1-form.

As a result, the fractional exterior derivative is defined as
\be \label{fed}
_ad^{\alpha}_x := [dx_m]^{\alpha} \, _a^CD^{\alpha}_{x_m}[x'_m] .
\ee
where
\[ [dx_m]^{\alpha}=sgn(dx_m) |dx_m|^{\alpha} \]
Then the fractional differential 1-form is
\be \label{fdof}
\omega(\alpha)=[dx_m]^{\alpha} \, F_m(x) . 
\ee
The exterior derivative of this form gives
\be \label{132}
_ad^{\alpha}_x \omega(\alpha)= [dx_m]^{\alpha} \wedge [dx_n]^{\alpha} 
\, _a^CD^{\alpha}_{x_n}[x']  F_m(x') . 
\ee

To prove the proposition (\ref{132}), we use the rule 
\[ D^{\alpha}_{x} (fg) =\sum^{\infty}_{s=0} 
\left(^{\alpha}_k \right) (\, _a^CD^{\alpha-s}_x f ) D^s_x g ,\]
and the relation \cite{KST} 
$D^s[x] (d x)^{\alpha} =0 \quad (s \ge 1)$,
for integer $s$, where
\[ \left(^{\alpha}_k \right)=
\frac{(-1)^{k-1} \alpha \Gamma(k-\alpha)}{\Gamma(1-\alpha) \Gamma(k+1)}. \]
For example, we have 
\[ d^{\alpha} \left[ [dx_m]^{\alpha} F_m \right]=
\sum^{\infty}_{s=0} [d x_n]^{\alpha} \wedge 
\left(^{\alpha}_k \right) (\, _a^CD^{\alpha-s}_{x_n} [{x'}_n] F_m(x') ) 
D^s [x_n] [d x_m]^{\alpha} =\]
\[ = [d x_n]^{\alpha} \wedge [d x_m]^{\alpha} 
\left(^{\alpha}_0 \right) \, ^CD^{\alpha}_{x_n}[{x'}_n] F_m(x')=
\left( ^CD^{\alpha}_{x_n}[{x'}_n] F_m(x') \right) \,
[d x_n]^{\alpha} \wedge [d x_m]^{\alpha} . \]

Using the equation (see Property 2.16 in \cite{KST})
\be
_aD^{\alpha}_x [x'] (x'-a)^{\beta}= \frac{\Gamma(\beta+1)}{\Gamma(\beta+1-\alpha)} 
(x-a)^{\beta-\alpha} ,
\ee
where $n-1<\alpha<n$, and $\beta>n-1$, and
\be
_aD^{\alpha}_x [x'] (x'-a)^{k}=0 \quad (k=0,1,2,...,n-1) ,
\ee
we obtain
\be
_ad^{\alpha}_x (x-a)^{\alpha}= [dx]^{\alpha} \ _a^CD^{\alpha}_{x}[x'] x' =
(dx)^{\alpha} \, \Gamma(\alpha+1)  \quad (x>a).
\ee
Then
\be
[dx]^{\alpha}= \frac{1}{\Gamma(\alpha+1) } \, _ad^{\alpha}_x (x-a)^{\alpha} ,
\ee
and the fractional exterior derivative (\ref{fed}) is presented as
\be
_ad^{\alpha}_x := \frac{1}{\Gamma(\alpha+1)} 
\, _ad^{\alpha}_x (x_m-a_m)^{\alpha} \, _a^CD^{\alpha}_{x_m}[x'_m] .
\ee
The fractional differential 1-form (\ref{fdof}) can be written as
\be 
\omega(\alpha)=\frac{1}{\Gamma(\alpha+1)} \, _ad^{\alpha}_x (x_m-a_m)^{\alpha} \, F_m(x) . 
\ee

{\bf Remark}. 
Using the suggested definition of fractional integrals and differential forms,
it is possible to define a fractional integration of 
$n$-form over the hypercube $[0,1]^n$.
Unfortunately, a generalization of this fractional integral,  
which uses the mapping $\phi$
of the region $W\subset \mathbb{R}^n$ into $[0,1]^n$, has a problem.
For the integer case, we use the equation
\be \label{dcf}
D^1_x f(\phi(x))=(D^1_{\phi} f) (D^1_x \phi) .
\ee
For the fractional case, the chain rule for differentiation
(the fractional derivative of composite functions)
is more complicated (see Section 2.7.3. \cite{Podlubny}).
As a result, a consistent definition of fractional integration of 
differential form for arbitrary manifolds is an open question.

\subsection{Differential vector operations through the differential forms}

To define a fractional divergence of the field ${\bf F}$, we can consider the 2-form
\be
\omega_2 = F_z dx \wedge dy +  F_y dz \wedge dx+ F_x dy \wedge dz .
\ee
Then the fractional exterior derivative of this form is
\be
d \omega_2 = (D_x F_x + D_y F_y+ D_z F_z ) dx \wedge dy \wedge dz=
div {\bf F} \, dx \wedge dy \wedge dz .
\ee

To define a fractional generalization of the curl operation for ${\bf F}$, 
we can use the 1-form 
\be
\omega_1 = F_x dx +  F_y dy + F_z dz .
\ee
Then the fractional exterior derivative of this 1-form is
\be
d \omega_1 = (D_x F_y - D_y F_x ) dx \wedge dy 
+ (D_y F_z - D_z F_y ) dy \wedge dz 
+ (D_x F_z - D_z F_x ) dx \wedge dz .
\ee

To define the fractional gradient, we consider the 0-form
\be
\omega_0 = f(x,y,z)
\ee
Then the  fractional exterior derivative of $f$ gives
\be
d \omega_0 = D_x f dx + D_y f dy + D_z f dz = 
\sum^3_{k=1}(grad f)_k \, dx^k.
\ee
It is not hard to obtain fractional generalizations of these definitions.


\section{Fractional nonlocal Maxwell's equations}

\subsection{Local Maxwell's equations}

The behavior of electric fields (${\bf E},{\bf D}$), 
magnetic fields (${\bf B}, {\bf H}$), charge density ($\rho(t,r)$),  
and current density (${\bf j}(t,r)$)
is described by the Maxwell's equations 

\be \label{ME1}
div \, {\bf D}(t,r)=\rho(t,r),
\ee
\be \label{ME2}
curl \, {\bf E}(t,r)=- \partial_t {\bf B}(t,r) ,
\ee
\be \label{ME3}
div \, {\bf B}(t,r)=0,
\ee
\be \label{ME4}
curl \, {\bf H}(t,r)={\bf j}(t,r)+ \partial_t {\bf D}(t,r) .
\ee
Here $r=(x,y,z)$ is a point of the domain $W$.
The densities $\rho(t,r)$ and ${\bf j}(t,r)$ 
describe an external sources. 
We assume that the external sources of electromagnetic field are given.

The relations between electric fields (${\bf E},{\bf D}$) 
for the medium can be realized by 
\be \label{D1}
{\bf D}(t,r)=\varepsilon_0 \int^{+\infty}_{-\infty} \varepsilon(r,r') {\bf E}(t,r') dr' ,
\ee
where $\varepsilon_0$ is the permittivity of free space.
Homogeneity in space gives $\varepsilon(r,r')=\varepsilon(r-r')$.
Equation (\ref{D1}) means that the displacement ${\bf D}$ is a convolution of 
the electric field ${\bf E}$ at other space points.
A local case corresponds to the Dirac delta-function permittivity 
$\varepsilon(r)=\varepsilon \delta(r)$. 
Then Eq. (\ref{D1}) gives ${\bf D}(t,r)=\varepsilon_0 \varepsilon {\bf E}(t,r)$.

Analogously, we have nonlocal equation for the magnetic fields (${\bf B}, {\bf H}$).

\subsection{Caputo derivative in electrodynamics}

Let us demonstrate a possible way of appearance of 
the Caputo derivative in the classical electrodynamics.
If we have
\be
{\bf D}(t,x) = \int^{+\infty}_{-\infty} \varepsilon (x-x') {\bf E}(t,x') dx' ,
\ee
then
\be
D^1_x {\bf D}(t,x)= 
\int^{+\infty}_{-\infty} [D^1_x \varepsilon (x-x')] {\bf E}(t,x') dx' =
-\int^{+\infty}_{-\infty} [D^1_{x'} \varepsilon (x-x')] {\bf E}(t,x') dx' .
\ee
Using the integration by parts, we get
\be \label{DeE}
D^1_x {\bf D}(t,x)= 
\int^{+\infty}_{-\infty} \varepsilon (x-x') D^1_{x'} {\bf E}(t,x')\, dx' .
\ee
Consider the kernel $\varepsilon (x-x')$ of integral (\ref{DeE}) 
in the interval $(0,x)$ such that
\be \label{Mtt} 
\varepsilon (x-x')=
\begin{cases} 
e (x-x') , & 0 < x' < x ;
\cr 
0, & x' > x, \quad \ x'< 0 ,
\end{cases}
\ee
with the power-like function 
\be \label{Mt1}
e(x-x') = \frac{1}{\Gamma(1-\alpha)} \frac{1}{(x-x')^{\alpha}} , 
\quad (0<\alpha <1) .
\ee
Then Eq. (\ref{DeE}) gives the relation
\be 
D^1_x {\bf D}(t,x)=\, _0^CD^{\alpha}_x {\bf E}(t,x), \quad (0<\alpha <1) 
\ee
with the Caputo fractional derivatives $\ _0^CD^{\alpha}_x$.

\subsection{Fractional nonlocal Maxwell's equations}

Fractional nonlocal differential Maxwell's equations have the form
\be \label{FME1}
Div^{\alpha_1}_W \, {\bf E}(t,r)= g_1 \rho(t,r),
\ee
\be \label{FME2}
Curl^{\alpha_2}_W  \, {\bf E}(t,r)=- \partial_t {\bf B}(t,r) ,
\ee
\be \label{FME3}
Div^{\alpha_3}_W  \, {\bf B}(t,r)=0,
\ee
\be \label{FME4}
g_2 Curl^{\alpha_4}_W  \, {\bf B}(t,r)=
{\bf j}(t,r)+ g^{-1}_3 \partial_t {\bf E}(t,r) ,
\ee
where $\alpha_s$, ($s=1,2,3,4$), can be integer or fractional.

Fractional integral Maxwell's equations, 
which use integrals of noninteger orders, 
have been suggested in \cite{T1} to describe 
fractional distributions of electric charges and currents.

In the general form, the fractional integral Maxwell's equations 
can be presented in the form
\be \label{FME1b}
\Bigl( {\bf I}^{\alpha_1}_{\partial W} , {\bf E}(t,r) \Bigr)= 
g_1 I^{\alpha_1}_W \rho(t,r),
\ee
\be \label{FME2b}
\Bigl( {\bf I}^{\alpha_2}_{\partial S} , {\bf E}(t,r)\Bigr) =
- \frac{d}{dt} \Bigl( {\bf I}^{\alpha_2}_S ,{\bf B}(t,r) \Bigr),
\ee
\be \label{FME3b}
\Bigl( {\bf I}^{\alpha_3}_{\partial W} , {\bf B}(t,r)\Bigr)=0,
\ee
\be \label{FME4b}
g_2  
\Bigl( {\bf I}^{\alpha_4}_{\partial S} , {\bf B}(t,r)\Bigr) =
\Bigl( {\bf I}^{\alpha_4}_S,  {\bf j}(t,r) \Bigr)+ 
g^{-1}_3 \frac{d}{dt}  \Bigl( {\bf I}^{\alpha_4}_S, {\bf E}(t,r) \Bigr) .
\ee

These fractional differential and integral equations can be used 
to describes an electromagnetic field 
of media that demonstrate fractional nonlocal properties.
The suggested equations can be considered as a special case 
of nonlocal electrodynamics (see \cite{NLE1,NLE2,NLE3,NLE4,NLE5}).

Fractional coordinate derivatives are connected 
with nonlocal properties of the media.
For example, a power-law long-range interaction in the 3-dimensional lattice
in the continuous limit can give a fractional equation \cite{JMP}.

\subsection{Fractional conservation law for electric charge}

Let us derive a conservation law equation for density of electric charge
in the region $W$ from the fractional nonlocal Maxwell's equations.

The time derivative of (\ref{FME1b}) is
\be \label{DdD}
Div^{\alpha_1}_W \, \partial_t {\bf E}(t,r)= g_1 \partial_t \rho(t,r) .
\ee
Substitution of (\ref{FME4}) into (\ref{DdD}) gives
\be
g_3 Div^{\alpha_1}_W \, \Bigl( g_2 Curl^{\alpha_4}_W  {\bf B}(t,r)-{\bf j}(t,r) \Bigr)= 
g_1 \partial_t \rho(t,r) .
\ee
If $\alpha_1=\alpha_4$, then
\be
Div^{\alpha_1}_W \, Curl^{\alpha_4}_W  \, {\bf B}(t,r)=0,
\ee
and we have the law
\be \label{cl1}
g_1 \partial_t \rho(t,r) + g_3 Div^{\alpha_1}_W \, {\bf j}(t,r) = 0.
\ee
This fractional equation is a differential form of charge conservation law
for fractional nonlocal electrodynamics.

If $\alpha_1=\alpha_4$, we can define the fractional integral characteristics such as 
\be
Q_W(t)= g_1 I^{\alpha_1}_W [x,y,z] \rho(t,x,y,z) ,
\ee
which can be called the total fractional nonlocal electric charge, and
\be
J_{\partial W}(t) = g_3 \Bigl( {\bf I}^{\alpha_1}_{\partial W} , {\bf j}\Bigr) =
g_3 \Bigl( I^{\alpha_1}_{\partial W}[y,z] j_x +
I^{\alpha_1}_{\partial W}[z,x] j_y +
I^{\alpha_1}_{\partial W}[x,y] j_z \Bigr) 
\ee
is a fractional nonlocal current.
Then the fractional nonlocal conservation law is
\be \label{cl2}
\frac{d}{dt}Q_W(t)+ J_{\partial W}(t)=0 .
\ee
This integral equation describes the conservation of the electric charge
in the nonlocal electrodynamics for the case $\alpha_1=\alpha_4$.

\subsection{Fractional waves}

Let us derive wave equations for electric and magnetic fields
in a region $W$ from the fractional nonlocal Maxwell's equations
with ${\bf j}=0$ and $\rho=0$.

The time derivative of Eq. (\ref{FME2}) is 
\be \label{BE}
\partial^2_t {\bf B}= - Curl^{\alpha_2}_W  \partial_t {\bf E}
\ee
Substitution of (\ref{FME4b}) and ${\bf j}=0$ into (\ref{BE}) gives
\be
\partial^2_t {\bf B}= - g_2 g_3 Curl^{\alpha_2}_W  \, Curl^{\alpha_4}_W  {\bf B}(t,r) .
\ee
Using (\ref{CC}) and (\ref{FME3}) for $\alpha_2=\alpha_3=\alpha_4$, we get
\be
\partial^2_t {\bf B}= g_2 g_3 (\, ^C{\bf D}^{\alpha}_W)^2 {\bf B} .
\ee 
As a result, we obtain
\be \label{FWE}
\partial^2_t {\bf B}-v^2 (\, ^C{\bf D}^{\alpha}_W)^2 {\bf B} =0 ,
\ee
where $v^2=g_2 g_3$.
This is the fractional wave equation for the magnetic field ${\bf B}$.
Analogously, Eqs. (\ref{FME2}) and (\ref{FME4b}) give 
the fractional wave equation for electric field
\be
\partial^2_t {\bf E}-v^2 (\, ^C{\bf D}^{\alpha}_W)^2 {\bf E} =0 .
\ee

The solution  ${\bf B}(t,r)$ of equation (\ref{FWE}) is a linear combination of 
the solutions ${\bf B}_{+}(t,r)$ and ${\bf B}_{-}(t,r)$ of the equations
\be \label{DA1}
\partial_t {\bf B}_{+}(t,r) - v\, ^C{\bf D}^{\alpha}_W {\bf B}_{+}(t,r)=0 , \
\ee
\be \label{DA2}
\partial_t {\bf B}_{-}(t,r) + v\, ^C{\bf D}^{\alpha}_W {\bf B}_{-}(t,r)=0 .
\ee
As a result, we get the fractional extension of D'Alembert expression 
that is considered in \cite{PV}.

For the boundary conditions
\be
\lim_{|t| \rightarrow \infty} {\bf B}(t,r)=0, \quad {\bf B}(t,0)={\bf G}(t) ,
\ee
the general solution of equations (\ref{DA1}) and (\ref{DA2}) is given \cite{KST} by
\be
B_{m \pm} (t,r) =\frac{1}{2\pi} \int^{+\infty}_{-\infty} d \omega 
E_{\alpha,1}[ \mp iv \omega x^{\alpha}_m]\, \tilde G_m (\omega) e^{-i \omega t} ,
\ee
where $\tilde G_m (\omega)={\cal F}[G_m(t)]$,  and
$E_{\alpha,\beta}[z]$ is the biparametric Mittag-Leffler function \cite{KST}.
Here $B_{\pm m}(t,r)$, and $G_m(t)$ are components 
of ${\bf B}_{\pm}(t,r)$ and ${\bf G}(t)$.

For one-dimensional case, $B_x(x,y,z,t)=u(x,t)$, $B_y=B_z=0$, and
we can consider the fractional partial differential equation 
\be \label{C1}
D^2_t u(x,t)-v^2\, _0D^{2\alpha}_x u(x,t)=0, \quad  x\in \mathbb{R}, \  x>0, \ v>0,
\ee
with the conditions
\be \label{C2}
D^k_x u(0,t)=f_k(t) ,
\ee
where $k=0$ for $0<\alpha\le 1/2$, and $k=1$ for $1/2<\alpha\le 1$.
If $0<2\alpha<2$ and $v>0$,
the system of equations (\ref{C1}), (\ref{C2}) is solvable (Theorem 6.3. of \cite{KST}), 
and the solution $u(x,t)$ is given by
\be
u(x,t)=\sum^{n-1}_{k=0} \int^{+\infty}_{-\infty}
G^{2\alpha}_k(y,t) f_k(y) dy , \quad (n-1 < \alpha \le n) ,
\ee
where
\be
G^{2\alpha}_k(x,t) =\frac{1}{2} v x^{k-\alpha} 
\phi(-\alpha,k+1-\alpha,v |t|x^{-\alpha}) .
\ee
Here $\phi(-\alpha,k+1-\alpha,v|t|x^{-\alpha})$ is the Wright function \cite{KST}.

Note that the solutions of equations as (\ref{DA1}) and (\ref{DA2})
are based primary on the use of Laplace
transforms for equations with the Caputo $\ _0^CD^{\alpha}_x$ derivatives.
This leaves certain problems \cite{KST} with 
the fractional derivatives $\ _a^CD^{\alpha}_x$ for $a\in \mathbb{R}$.


\section{Conclusion}

Let us note some 
possible extensions of the fractional vector calculus. \\

(1) It is very important to prove the suggested fractional integral theorems 
for a general form of domains and boundaries. 

(2) It is interesting to generalize the formulations of fractional integral 
theorems for $\alpha>1$. 

(3) A proof of fractional theorems for differential forms can be interesting 
to formulate a fractional generalization of differential geometry. \\

In the fundamental theorem of fractional calculus (FTFC) 
we use the Riemann-Liouville integration and the Caputo differentiation.
The main property is that the Caputo fractional derivative  
provides us an operation inverse to the Riemann-Liouville 
fractional integration from the left. 
Note that a fractional generalization of 
the differential vector operations and 
the integral theorems for the fractional integro-differentiation
of Riesz, Grunvald-Letnikov, Weyl, Nishimoto is an open problem.

There are the following possible applications of 
the fractional variational calculus (FVC). \\

(a) A fractional nonlocal electrodynamics that is characterized 
by the power law non-locality can be formulated by using the FVC. 

(b) Nonlocal properties in classical dynamics can be described by the FVC
and by possible fractional generalizations of symplectic geometry
and Poisson algebra. In general, fractional differential forms
and fractional integral theorems for these forms can be used 
to describe classical dynamics. 

(c) A possible dynamics of fractional gradient and Hamiltonian dynamical systems
can be described by the FVC. 

(d) The continuum mechanics of fluids and solids
with nonlocal properties (with a nonlocal interaction of medium particles) 
can be described by the FVC. \\

The fractional derivatives in equations can be connected with 
a long-range power-law interaction of the systems \cite{LZ,TZ3,JMP}. 
The nonlocal properties of electrodynamics can be considered \cite{JPC2}
as a result of dipole-dipole interactions with a fractional power-law screening 
that is connected with the integro-differentiation of non-integer order.
For noninteger derivatives with respect to coordinates, 
we have the power-like tails as the important property 
of the solutions of the fractional equations.



\end{document}